\documentclass[twocolumn]{aastex63}
\newcommand{\roma}[1]{\uppercase\expandafter{\romannumeral#1}}
\newcommand{\speed}[1]{#1 km~s${}^{-1}$}

\newcommand{\nfig}[1]{Figure~\ref{#1}}

\usepackage{amsmath}
\usepackage{subfigure}
\usepackage{hyperref}%

\usepackage{graphicx}
\usepackage{natbib}
\usepackage{amssymb,txfonts}
\usepackage{multirow}
\usepackage{array}
\usepackage{sidecap}
\usepackage{hyperref}
\usepackage{epstopdf}
\usepackage{footnote}
\usepackage{tabularx}
\usepackage{booktabs}
\shorttitle{Observations of the sympathetic coronal jets}
\shortauthors{Tang et al.}

\begin{document}
\title{Sympathetic Standard and Blowout Coronal Jets Observed in a Polar Coronal Hole}
\correspondingauthor{Yuandeng Shen}
\email{ydshen@ynao.ac.cn}
\author{Zehao Tang}
\affiliation{Yunnan Observatories, Chinese Academy of Sciences,  Kunming, 650216, China}
\affiliation{State Key Laboratory of Space Weather, Chinese Academy of Sciences, Beijing 100190, China}
\affiliation{University of Chinese Academy of Sciences, Beijing, China}
\author[0000-0001-9493-4418]{Yuandeng Shen}
\affiliation{Yunnan Observatories, Chinese Academy of Sciences,  Kunming, 650216, China}
\affiliation{State Key Laboratory of Space Weather, Chinese Academy of Sciences, Beijing 100190, China}
\author{Xinping Zhou}
\affiliation{Yunnan Observatories, Chinese Academy of Sciences,  Kunming, 650216, China}
\affiliation{University of Chinese Academy of Sciences, Beijing, China}
\author{Yadan Duan}
\affiliation{Yunnan Observatories, Chinese Academy of Sciences,  Kunming, 650216, China}
\affiliation{University of Chinese Academy of Sciences, Beijing, China}
\author{Chengrui Zhou}
\affiliation{Yunnan Observatories, Chinese Academy of Sciences,  Kunming, 650216, China}
\affiliation{University of Chinese Academy of Sciences, Beijing, China}
\author{Song Tan}
\affiliation{Yunnan Observatories, Chinese Academy of Sciences,  Kunming, 650216, China}
\affiliation{University of Chinese Academy of Sciences, Beijing, China}
\author{Elmhamdi Abouazza}
\affiliation{Physics and Astronomy Department, College of Science, King Saud University, Riyadh, Saudi Arabia}

\begin{abstract}
We present the sympathetic eruption of a standard and a blowout coronal jets originating from two adjacent coronal bright points (CBP1 and CBP2) in a polar coronal hole, using soft X-ray and extreme ultraviolet observations respectively taken by the {\em Hinode} and the {\em Solar Dynamic Observatory}. In the event, a collimated jet with obvious westward lateral motion firstly launched from CBP1, during which a small bright point appeared around CBP1's east end, and magnetic flux cancellation was observed within the eruption source region. Based on these characteristics, we interpret the observed jet as a standard jet associated with photosperic magnetic flux cancellation. About 15 minutes later, the westward moving jet spire interacted with CBP2 and resulted in magnetic reconnection between them, which caused the formation of the second jet above CBP2 and the appearance of a bright loop system in-between the two CBPs. In addition, we observed the writhing, kinking, and violent eruption of a small kink structure close to CBP2's west end but inside the jet-base, which made the second jet brighter and broader than the first one. These features suggest that the second jet should be a blowout jet triggered by the magnetic reconnection between CBP2 and the spire of the first jet. We conclude that the two successive jets were physically connected to each other rather than  a temporal coincidence, and this observation also suggests that coronal jets can be triggered by external eruptions or disturbances, besides internal magnetic activities or magnetohydrodynamic instabilities.
\end{abstract}

\section{Introduction}
Coronal jets, an ubiquitous phenomenon in the solar atmosphere, are heated plasmas flows moving along magnetic field lines showing as collimated or two-sided ejections \citep[e.g.,][]{1992PASJ...44L.173S,1995Natur.375...42Y,2011ApJ...735L..43S,2019ApJ...883..104S,2019ApJ...887..220Y,2021RSPSA.47700217S,2018ApJ...861..108Z,2016ApJ...830...60H}; they can occur in active regions, quiet-sun regions, and corona holes; and their lengths, widths, lifetimes, and velocities are in the ranges of a few $\times10^{4}$ to 4 $\times10^{5}$ km, 5$\times10^{3}$ to $10^{5}$ km, a few minutes to over ten hours and $10$ to \speed{$10^{3}$}, respectively \citep{1996PASJ...48..123S,1998SoPh..178..379S,2021RSPSA.47700217S}. Previous observational and numerical studies have revealed that the basic physical process in coronal jets is the magnetic reconnection \citep[e.g.,][]{1996ApJ...464.1016C,2011ApJ...735L..43S,2012ApJ...745..164S,2017ApJ...845...94T,2019ApJ...871..220S,2021RSPSA.47700217S}, and photospheric magnetic flux emergence and cancellation is the most common triggering reason \citep[e.g.,][]{2007A&A...469..331J,2011RAA....11.1229Y,2011ApJ...728..103L,2012ApJ...745..164S,2017ApJ...851...67S,2016ApJ...817...39L,2016ApJ...832L...7P,2018ApJ...864...68S,2018ApJ...853..189P,2009SoPh..255...79C,2012A&A...539A...7L,2011ApJ...738L..20H,2015Ap&SS.359...44L}. Recent high spatiotemporal resolution observations have revealed that many coronal jets are caused by mini-filament eruptions, thus that they may represent the miniature version of large-scale, energetic solar eruptions such as filament eruptions and coronal mass ejections (CMEs) \citep[e.g.,][]{2010ApJ...720..757M,2012ApJ...745..164S,2017ApJ...851...67S,2019ApJ...885L..11S,2015Natur.523..437S,2019ApJ...887..239Y,2020ApJ...897..113H,2020ApJ...902....8C,2021arXiv210106629Z}. In addition, coronal jets are  important for triggering large-scale solar phenomena such as coronal waves and CMEs \citep[e.g.,][]{2012ApJ...745..164S,2018ApJ...861..105S,2018ApJ...860L...8S,2016ApJ...823..129A,2016ApJ...822L..23P,2018ApJ...869...39M,2018MNRAS.480L..63S,2019ApJ...881..132D}, as well as for the heating of coronal plasma and the acceleration of solar wind \citep[e.g.,][]{2007Sci...318.1591S,2014Sci...346A.315T,2019ApJ...884L..51Y,2021RSPSA.47700217S}.

For a long time, it was equivocal that solar eruptions occurring at different sites during a relatively short time interval are physically connected or just a coincidence of time. Such kind of successive eruptions have been extensively studied, and those showing internal physical connections were called sympathetic eruptions \citep{2002ApJ...574..434M,2003ApJ...588.1176M,2012ApJ...750...12S}. Some studies revealed that the causal links between sympathetic solar eruptions are often of a magnetic nature. For example, sympathetic filament eruptions are often caused by magnetic reconnection around separatrices, separators, and quasi-separatrix layers \citep[e.g.,][]{2011JGRA..116.4108S,2011ApJ...739L..63T,2012ApJ...759...70T,2013ApJ...764...87L,2016ApJ...820..126J,2018ApJ...869..177W}, magnetic loops expansion due to impingement of external disturbances \citep[e.g.,][]{2011ApJ...738..179J,2012ApJ...745....9Y},  magnetic implosion mechanism within the framework of magnetic breakout configuration \citep{2012ApJ...750...12S,2018ApJ...864...68S}, and large-scale coronal waves \citep{2014ApJ...786..151S}. So far, sympathetic solar eruptions including flares \citep{2001ApJ...559.1171W,2002ApJ...574..434M}, filaments \citep[e.g.,][]{2011ApJ...738..179J,2012ApJ...750...12S,2020ApJ...892...79S,2020A&A...640A.101H}, and CMEs \citep[e.g.,][]{2003ApJ...588.1176M,2008ApJ...677..699J,2012ApJ...750...12S} have been documented in the historical literature. However, to the best of our knowledge, there is still no report on sympathetic coronal jets. 

In this letter, for the first time, we report the observation of a sympathetic jet event occurred on 2019 March 31, in which two successive coronal jets were observed due to the eruptions of two adjacent coronal bright points (CBPs) in the south polar coronal hole. The present study focuses on the eruption mechanism of the two coronal jets, and our analysis results suggest that they were physically connected with each other, i.e., the lateral sweeping motion of the first jet resulted in the onset of the second one. Since coronal jets are usually initiated or triggered by internal activities such as photospheric magnetic flux cancellations, or magnetohydrodynamics instabilities of mini-filaments or flux ropes, the present study showed evidence that coronal jets can also be launched by external eruptions or disturbances. The used observations are described in Section 2; results are presented in Section 3; discussions and conclusions are given in Section 4.

\begin{figure*}[tbh]
\epsscale{0.95}
\figurenum{1}
\plotone{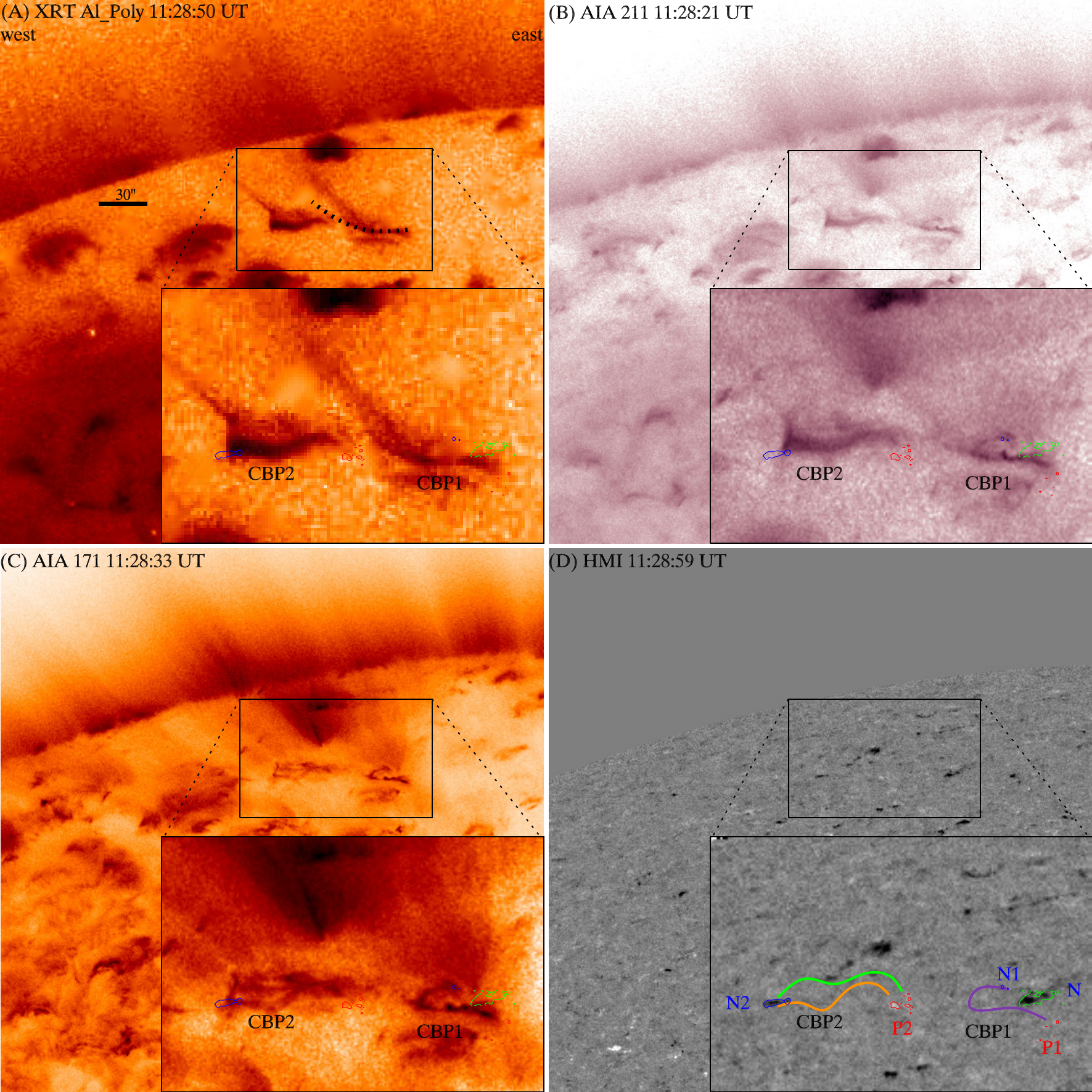}
\caption{Overview of the event. Panels (A)--(D) are XRT Al\_Poly, AIA 211\AA, AIA 171 \AA\ and HMI LOS snapshots, respectively. The field of view (FOV) of each frame is 300\arcsec\ $\times$ 400\arcsec. The FOV of each inset is 120\arcsec\ $\times$ 80\arcsec. The red and blue contours in each frame represent positive and negative magnetic polarities, and the green one outlines the negative magnetic polarity located on the east side of CBP1. The two ends of CBP1 (CBP2) were labeled with P1(P2) and N1(N2) in the magnetogram, and the polarity on the east side of CBP1 is labeled with N. In panel (D), CBP1 is outlined with a purple curve, and CBP2 is represented by the green and orange curves. 
\label{fig1}}
\end{figure*}

\section{Observations} 
The present event was recorded simultaneously by the X-ray Telescope \cite[XRT;][]{2007SoPh..243...63G} onboard the {\em Hinode}, the Atmospheric Imaging Assembly \cite[AIA;][]{2012SoPh..275...17L} and Helioseismic and Magnetic Imager \cite[HMI;][]{2012SoPh..275..207S} onboard the {\em Solar Dynamic Observatory} ({\em SDO}). During our observing time interval, {\em Hinode}/XRT provided Al\_Poly and Al\_Mesh  soft X-ray images, and the pixel size and cadence of these images are of 1\arcsec.02 and 37 seconds, respectively. {\em SDO}/AIA provided continuous full-disk observations of the solar chromosphere and corona in seven extreme ultraviolet (EUV) channels, spanning a temperature range from approximately $2 \times 10^{4}$ Kelvin to in excess of $2 \times 10^{7}$ Kelvin. Here, we only use the 171 \AA\ (Fe \roma{9}; characteristic temperature: $0.6 \times 10^{6}$ K) and 211 \AA\ (Fe \roma{14}; characteristic temperature: $2 \times 10^{6}$ K) images, since the evolution processes are similar or not observed in other AIA channels. The time cadence and pixel size of AIA images are 12 second and 0\arcsec.6, respectively. The line-of-sight (LOS) magnetograms taken by the HMI are used to analyze the magnetic flux variations within the eruption source region, whose cadence, pixel size, and measuring precision are of 45 second, 0\arcsec.5, and 10 Gauss, respectively.

\section{Results} 
The present event occurred in the south polar coronal hole on 2019 March 31. An overview of the eruption source region is presented in \nfig{fig1}, in which the images were rotated 180 degrees clockwise so that the eruption source region looks like in the north polar coronal pole (the same in the following figures). To better show of the imaging features, the soft X-ray and EUV images are displayed as negative images so that the original bright (dark) features manifest as black (white) structures. In the XRT Al\_Poly, AIA 171 \AA\ and 211 \AA\ images, one can identified many CBPs in the polar coronal hole as well as some coronal jets originating from these CBPs (see \nfig{fig1} and the animation in the accompanying online material). We focus on the two successive jets originated from two adjacent CBPs as indicated by the small black box in \nfig{fig1}, and this region was magnified and plotted as an inset in each panel. It is measured that the projection lengths of the two CBPs were less than 30\arcsec (see the black bar in \nfig{fig1}(A)).

CBP1 showed a lying J-shaped structure consisting of simple potential loops, while CBP2 looks like a mini-sigmoid structure composed of two evident sigmoidal strands (see \nfig{fig1}(A)--(C)). A line-of-sight (LOS) magnetogram is displayed in \nfig{fig1}(D) to show the magnetic field in the eruption source region, in which the white (black) patches represent positive (negative) polarities. The purple curve outlines the shape of CBP1 which connects opposite magnetic polarities of P1 (positive) and N1 (negative), while the green and orange curves outline the two strands of CBP2 linking opposite polarities of P2 (positive) to those of N2 (negative). It is noted that a negative magnetic polarity (N) was located in-between P1 and N1. The overall eruption process can be divided into three stages: the formation of the first jet (jet1: 11:10 UT -- 11:30 UT), the interaction of jet1 with CBP2 (11:30 UT -- 11:37 UT), and the formation of the second jet (jet2: 11:37 UT -- 12:00 UT). In the following subsections, we discuss the eruption characteristics in association with those stages in more detail.

\begin{figure*}[t]
\figurenum{2}
\epsscale{0.95}
\plotone{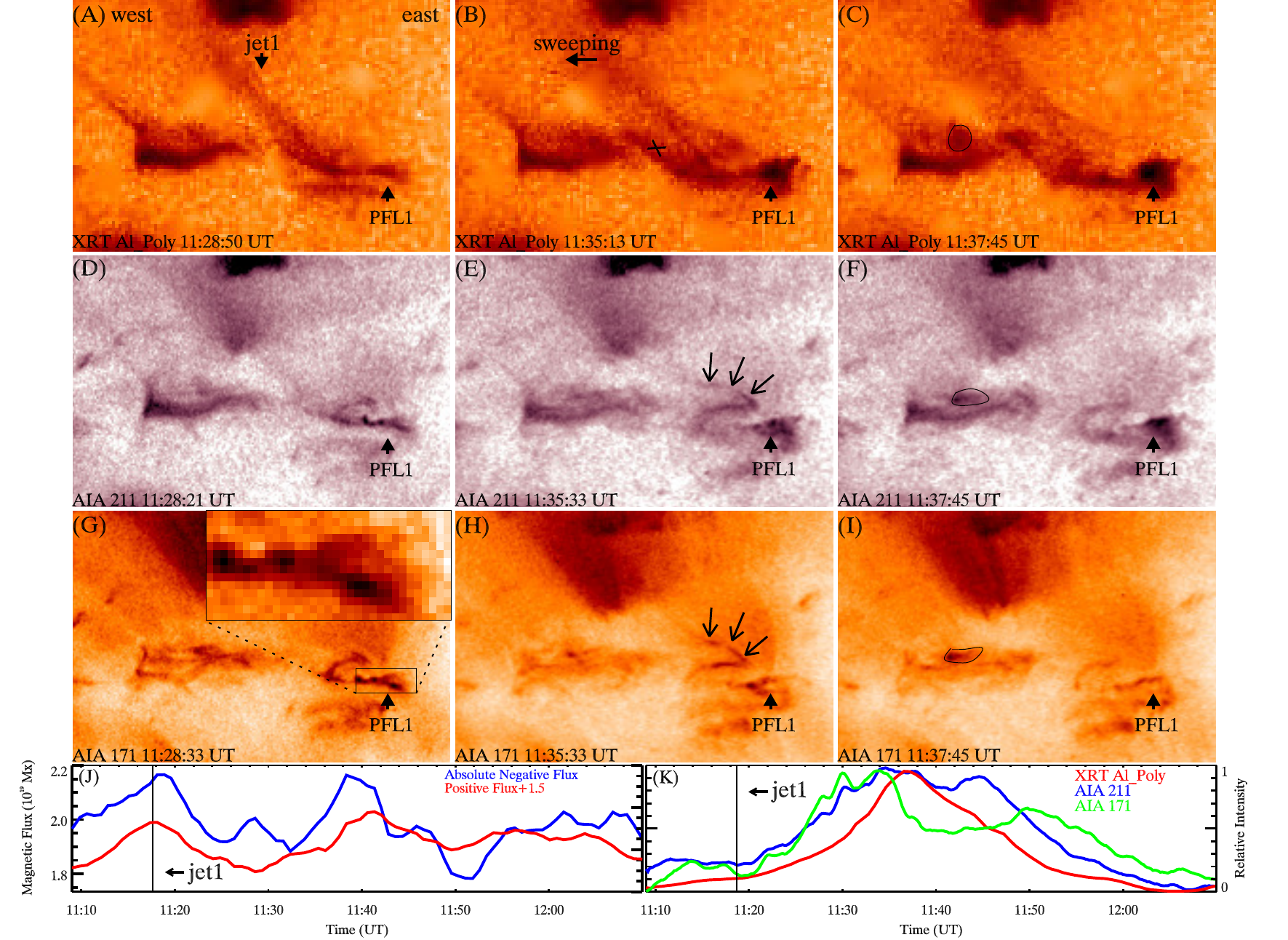}
\caption{Eruption of Jet1. Panels (A)--(C), (D)--(F) and (G)--(I) are XRT Al\_Poly, AIA 211 \AA\, and AIA 171 \AA\ time sequence images (time runs left to right). The downward arrow in panel (A) indicates the spire of jet1, while the upward arrows in panels (A)--(I) indicate PFL1. The arrow in panel (B) indicates the westward moving spire of jet1, and the black ``X" symbol marks the interaction site between the spire of jet1 and CBP2. The arrows in panels (E) and (H) indicate jet1 observed in AIA 211 \AA\ and 171 \AA\ images. The black contours in panels (C), (F), and (I) outline the westward moving plasmoid along the main axis of CBP2. ``PFL1" in panel (A)--(I) represents the post-flare-loop. Panel (J) and (K) show the plot of the magnetic flux variation within the eruption source region of jet1 and the corresponding intensity lightcureves at different wavelength bands, and the vertical black line in the two panels indicate the start time of jet1. (An animation of this figure named ``animation.mpeg" with 12s cadence is available, whose ratio and layout are same to this figure (exclude panels (J) and (K)), and duration is from 11:09 UT to 12:13 UT, covering the period that this picture illustrates.) 	
\label{fig2}}

\end{figure*}

\subsection{Formation of Jet1 and Its Interaction with CBP2}
\nfig{fig2} shows the formation of jet1 and its interaction with CBP2, in which the first, second, and third rows are XRT Al\_Poly, AIA 211 \AA, and AIA 171 \AA\ time sequence images, respectively. In the bottom row, the left panel shows the magnetic flux variations (measured from the HMI LOS magnetograms) within the eruption source region of CBP1, while the right panel shows the corresponding relative intensity curve (measured from XRT Al\_Poly, AIA 211 \AA, and AIA 171 \AA\ images) within the same region.

The left column of \nfig{fig2} shows CBP1 and jet1. The rising of the lying J-shaped CBP1 caused the collimated jet1, and the jet spire showed an obvious westward lateral sweeping motion as indicated by the arrows in \nfig{fig2}(A) and (B). In the meantime, a bright patch around the eastern end of CBP1 appeared, which has a typical inverted-Y (or cusp) shape and can be regarded as the post-flare-loop (PFL1) caused by the reconnection between CBP1 and its ambient open field lines (see the upward arrows in \nfig{fig2}(A)--(I) and the inset in \nfig{fig2}(G)). An overall consideration of the appearance locations of the jet, PFL1, and the magnetic polarities as shown in \nfig{fig1}(D), the magnetic reconnection should occur between CBP1 and the open field lines rooted in the negative magnetic region N, and this can also explain the westward lateral motion of the jet spire. From the EUV and soft X-ray intensity lightcurves in \nfig{fig2}(K), one can identified that the start of jet1 was about 11:18 UT (see the vertical black line in each panel). For the magnetic flux variations (\nfig{fig2}(J)), both the positive and absolute value of the negative magnetic fluxes showed obvious rising (declination) before (after) the start of the jet. Such a variation pattern of the magnetic fluxes suggests the first emerging and then cancellation of the opposite magnetic polarities in the eruption source region. Therefore, the jet was probably due to the magnetic reconnection between the emerging (rising) of CBP1 and the ambient pre-existing open field lines, while the flux cancellation probably manifests the submerging of the PFL1 connecting P1 and N. The above observational characteristics suggest that jet1 should be a standard jet \citep[e.g.,][]{1995Natur.375...42Y, 2010ApJ...720..757M}.

In a few minutes after the start of jet1, CBP1 expanded rapidly and the jet spire swept towards CBP2, which finally interacted with the eastern part of CBP2 (the interaction site is indicated by the ``X" symbol (see \nfig{fig2}(B))). This interaction lasted for a few minutes (less than 5 minutes) and can be identified in the XRT Al\_Poly images. In the AIA EUV observations, only the jet base can be resolved (see black arrows in \nfig{fig2}(E) and (H)). At 11:37:45 UT, the jet can be observed in the XRT Al\_Poly images (\nfig{fig2}(C)) but not in AIA images (\nfig{fig2}(F) and (I)). In the meantime, CBP2 lost its stability gradually, and a westward moving plasmoid along CBP2 can be observed (see the black circle in the right column of \nfig{fig2}). This feature may represent the accelerated plasma by the magnetic tension of the newly forming open field lines during the reconnection between the spire of jet1 and CBP2. Such a moving plasmoid-like feature was also previously evidenced to be a signature of the reconnection both in observational and numerical simulation studies  \citep{2013ApJ...764...87L,2016ApJ...827....4W,2019ApJ...885L..15K}.

\subsection{Formation of Jet2 and Its Fine Structure}
\begin{figure*}[t]
\epsscale{0.95}
\figurenum{3}
\plotone{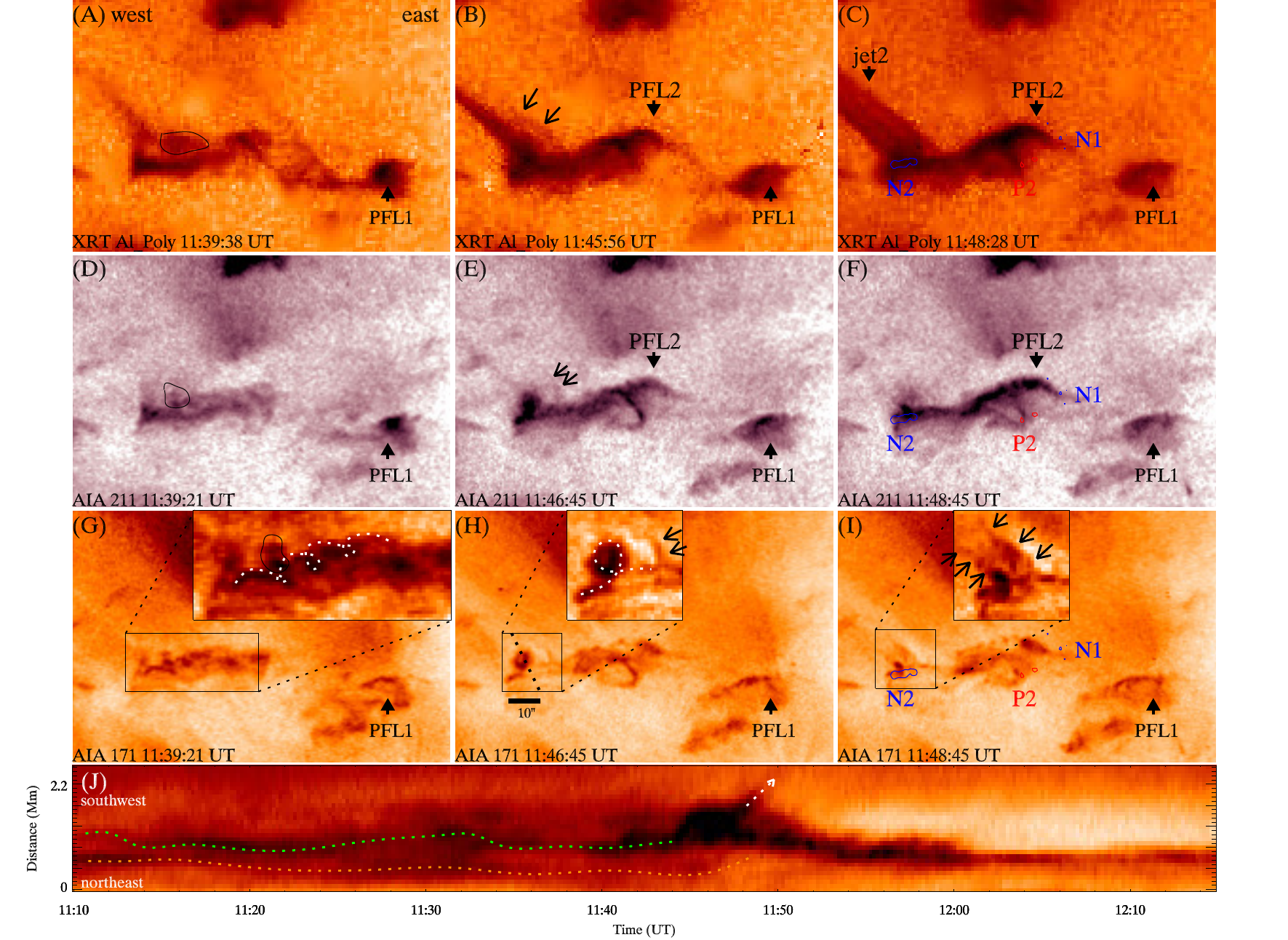}
\caption{Eruption of jet2. Panels (A)--(C), (D)--(F) and (G)--(I) are XRT Al\_Poly, AIA 211 \AA\ and AIA 171 \AA\ time sequence images (time runs left to right). In panels (A), (D) and (G), the westward moving plasmoid is outlined by the black contour. A close-up view of the solenoid-like structure is plotted as an inset in panel (G), and the structure is highlighted with a white dotted curve. In panels (B), (E) and (H), the newly formed open reconnected line (spire of jet2) is indicated by the black arrows. A close-up view of the small kink feature and newly formed spire is plotted as an inset and two arrows in panel (H), and the structure  is highlighted by white dotted curve. The left and right arrows in panel (C) indicate the spire of jet2 and the corresponding PFL2, while the labels N1, P2, and N2 in panels (C), (F), and (I) indicate the magnetic polarities as shown in panel \nfig{fig1}(D). The inset in panel (I) shows the fine structure of jet2, and the arrows indicate the edges of the jet spire. Panel (J) is a TDS plot made along the black dotted line as shown in panel (H), in which the two colored dotted curves trace the evolution of CBP2's two strands, and the white arrow indicates the eruption of the small kink structure. (An animation of this figure named ``animation.mpeg" with 12s cadence is available, whose ratio and layout are same to this figure (exclude panel (J)), and duration is from 11:09 UT to 12:13 UT, covering the period that this picture illustrates.)
\label{fig3}}
\end{figure*}

The launching process of jet2 and its fine structure are displayed in \nfig{fig3}, based on the time sequence images of XRT Al\_Poly (top row), AIA 211 \AA\ (second row), and AIA 171 \AA\ (third row) observations. One can distinguish that CBP2 was composed of two strand of loops (see \nfig{fig3}(D)), which expanded obviously and the small plasmoid kept its westward moving along the south loop of CBP2 (see the black circles in \nfig{fig3}(A), (D) and (G)). During this period, the south loop evolved into a solenoid-like structure as depicted by the white dotted curve in the inset in \nfig{fig3}(G). The formation of this special structure may due to the relaxing of the loop caused by the magnetic reconnection between the spire of jet1 and the eastern part of CBP2. Here, the south loop in CBP2 was initially a closed loop structure rooted on the solar surface; however, its east part was broken and became open loop due to the magnetic reconnection caused by the interaction. 

Specially, we identified the formation of a small bright kink structure inside the base of the second jet about 7 minutes after the start of the reconnection between jet1 and CBP2, which showed an inverted $\gamma$ shape and was close to the west end of CBP2 (see \nfig{fig3}(B), (E) and (H), and the white dotted curve in the inset in \nfig{fig3}(H)). Through checking the XRT and AIA time sequence images carefully (see also the online animation), it can be distinguished that the birth of the small kink structure was owning to one of the evolving knots on the solenoid-like loop. The kink structure was small and short-lived, whose lifetime and diameter were about 5 minutes and 10\arcsec\, respectively. During this time interval, the westward moving open section of the newly formed reconnected field line can be observed from the middle part of CBP2 (see the black arrows in \nfig{fig3}(B), (E) and the inset in panel (H)). In addition, a group of hot and bright loops connecting opposite polarities of N1 and P2 appeared and expanded significantly around the east end of CBP2 (see \nfig{fig3}(B), (C), (E) and (F)). Considering the connectivity of the coronal loops and the photospheric magnetic morphology, this newly formed hot loop system can be regarded as the post-flare-loop (PFL2) produced by the magnetic reconnection between CBP2 and the spire of jet1. 

The kink structure finally erupted southwesterly at about 11:48:00 UT, coinciding with the arrival of the westward moving loop getting close to the west end of CBP2 (see the black arrows in \nfig{fig3}(H)). The eruption of the kink structure was probably due to the internal reconnection between its two crossed legs, which can be confirmed by the appearance of two features: a cusp-like bright loop and a circular blob below and above the crossing point of the kink structure, respectively. The upward erupting blob merging with the westward moving loop and the pre-existing open loops around the west end of CBP2 formed the spire of jet2, while the cusp-like bright loop structure formed the jet base. Till to this time, jet2 was finally formed and it had a bigger size and brighter brightness than jet1. During the formation of jet2, CBP2 lost its double-loop structure during a short time interval. We create a time-distance stack plot to analyze the evolution of CBP2 and the eruption of the small kink structure (see \nfig{fig3}(J)). Here, the time-distance stack plot was made by composing the time-sequence of the intensity profiles of 171 \AA\ images along the black dotted line crossing this small kink in \nfig{fig3}(H), in which the abscissa and ordinate correspond to time and distance, respectively. The two loops and the eruption of the small kink structure are indicated by the two colored dotted curves and a white dotted arrow, respectively. It is measured that the upward erupting blob from the kink structure was at a speed of about \speed{130}. After the eruption of the kink structure, the two loops of CBP2 merged into one. 

Taking into account the overall evolution characteristics of CBP2, the eruption of the small kink structure, and the westward moving loop, we can conclude that jet2 was a blowout jet involving the eruption of non-potential magnetic field within its base \citep{2010ApJ...720..757M}, and it was trigged by the magnetic reconnection between the spire of jet1 and CBP2, owning to their interaction.

\subsection{Temperature Evolution and Physical Relation between the Two CBPs}
\begin{figure*}
\epsscale{0.95}
\figurenum{4}
\plotone{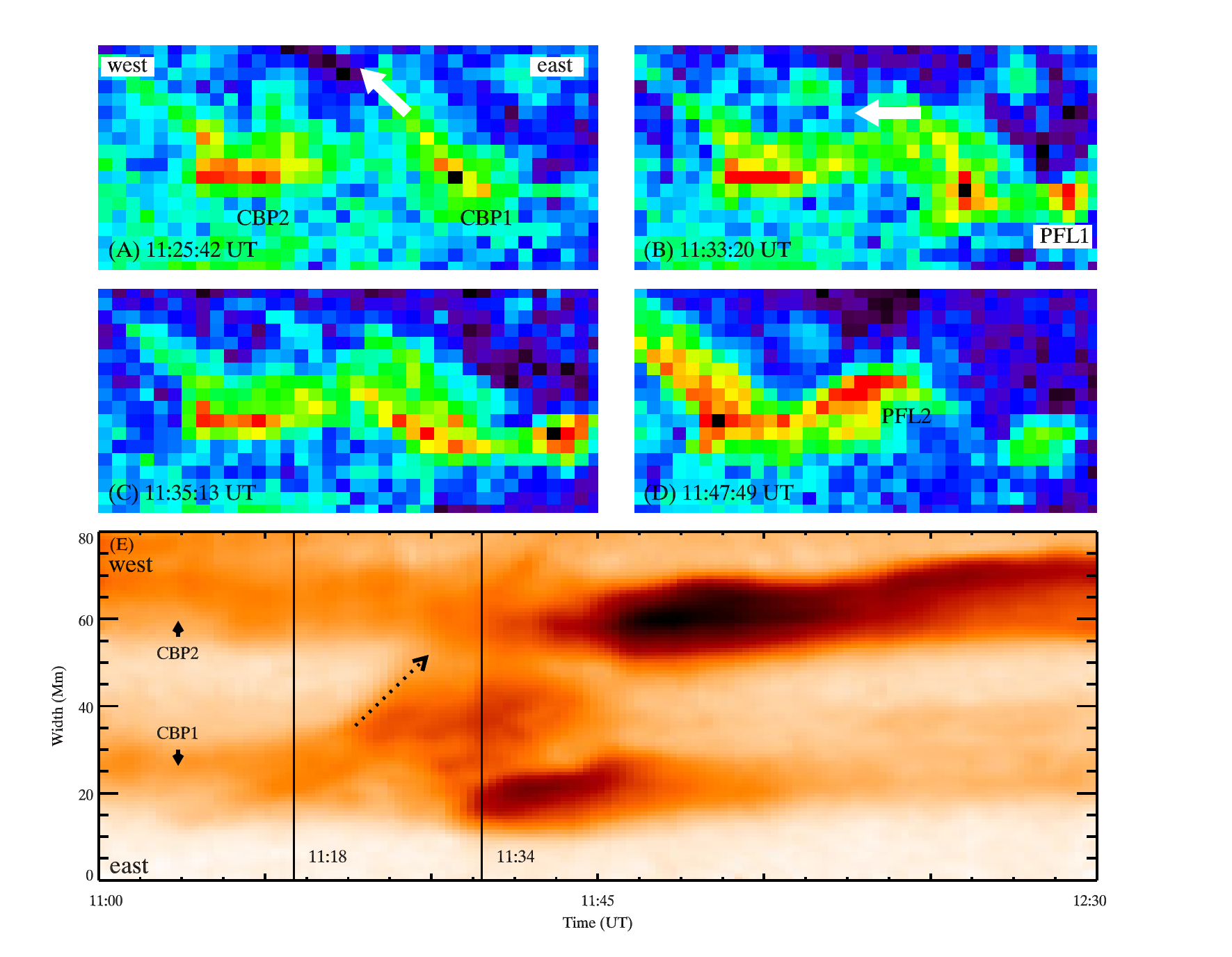}
\caption{
Panels (A)--(D) are temperature maps. The white arrow in panel (A) shows the direction of jet1, and the one in panel (B) indicates its westward sweeping motion. The locations of CBP1, CBP2, PFL1, and PFL2 are all marked in the maps. Panel (E) is a TDS plot made along the black dotted curve as shown in \nfig{fig1}(A), in which CBP1 and CBP2 indicate position of the two CBPs, and their eruption start times are indicated by the two vertical black lines. The westward sweeping motion of the spire of jet1 is indicated by the black arrow in panel (E).
\label{fig4}}
\end{figure*}

The temperature evolution of the event and the physical relationship between jet1 and jet2 are analyzed, and the results are presented in \nfig{fig4}. To obtain a temperature map with {\em Hinode} XRT data, one needs images at least from two different filters, and then processes them with the ``xrt\_teem.pro" procedure available in the SolarSoftWare (SSW) package. This procedure generates  temperature maps through solving the filter-ratio equation involving parameters of temperature ``T" and the ratio of two different filters' temperature responses \cite{2011SoPh..269..169N}. Here, we used the XRT Al\_Poly and Al\_Mesh data to generate the temperature maps, in which the redder (bluer) color corresponds to higher (lower) temperature (see \nfig{fig4}(A)--(D)). For a coronal jet, the spire, footpoint, and the brightening patch next to the footpoint should be high temperature regions, due to the deposition of hot plasma and energetic particles produced by the reconnection that triggered the jet \citep{1995Natur.375...42Y,2010ApJ...720..757M,2012ApJ...745..164S}. One can see that the temperature maps well reflected the evolution characteristics of the sympathetic jets. The spire of jet1 can be identified as a red elongated feature at the 11:25:42 UT (see the white arrow in \nfig{fig4}(A)). Several minutes later, the brightening feature appeared at the east side of jet1's footpoint. In the meantime, jet1 showed a strong westward lateral sweeping motion as indicated by the white arrow in \nfig{fig4}(B). Then, the spire of jet1 interacted with eastern part of CBP2 (see \nfig{fig4}(C)). After that, CBP2 was activated and produced jet2 around the its west end, and also PFL2 around the east end of CBP2 (see \nfig{fig4}(D)).

The interaction between the spire of jet1 and CBP2 is also probed by using a time-distance stack plot (\nfig{fig4}(E)) made by composing the time-sequence of the intensity profiles of XRT Al\_Poly images along the black dotted curve in \nfig{fig1}(A). In this time-distance stack plot, the two CBPs show as two horizontal thick lanes before the event. From about 11:18 UT when jet1 started, one can clearly identify the sweeping process of the spire of jet1 passing through the gap between CBP1 and CBP2 (see the black arrow in \nfig{fig4}(E)), and the westward lateral sweeping speed is measured to be about \speed{19}. The spire of jet1 interacted with CBP2 at about 11:34 UT, after that CBP2 was activated and brightened significantly. By examining the spatial position along the path used to make the time-distance stack plot, the brightest parts along the lanes of CBP1 and CBP2 are indeed the locations of PFL1 and PFL2, respectively.  

\subsection{Physical Interpretation}    
\begin{figure*}[t]
\epsscale{0.95}
\figurenum{5}
\plotone{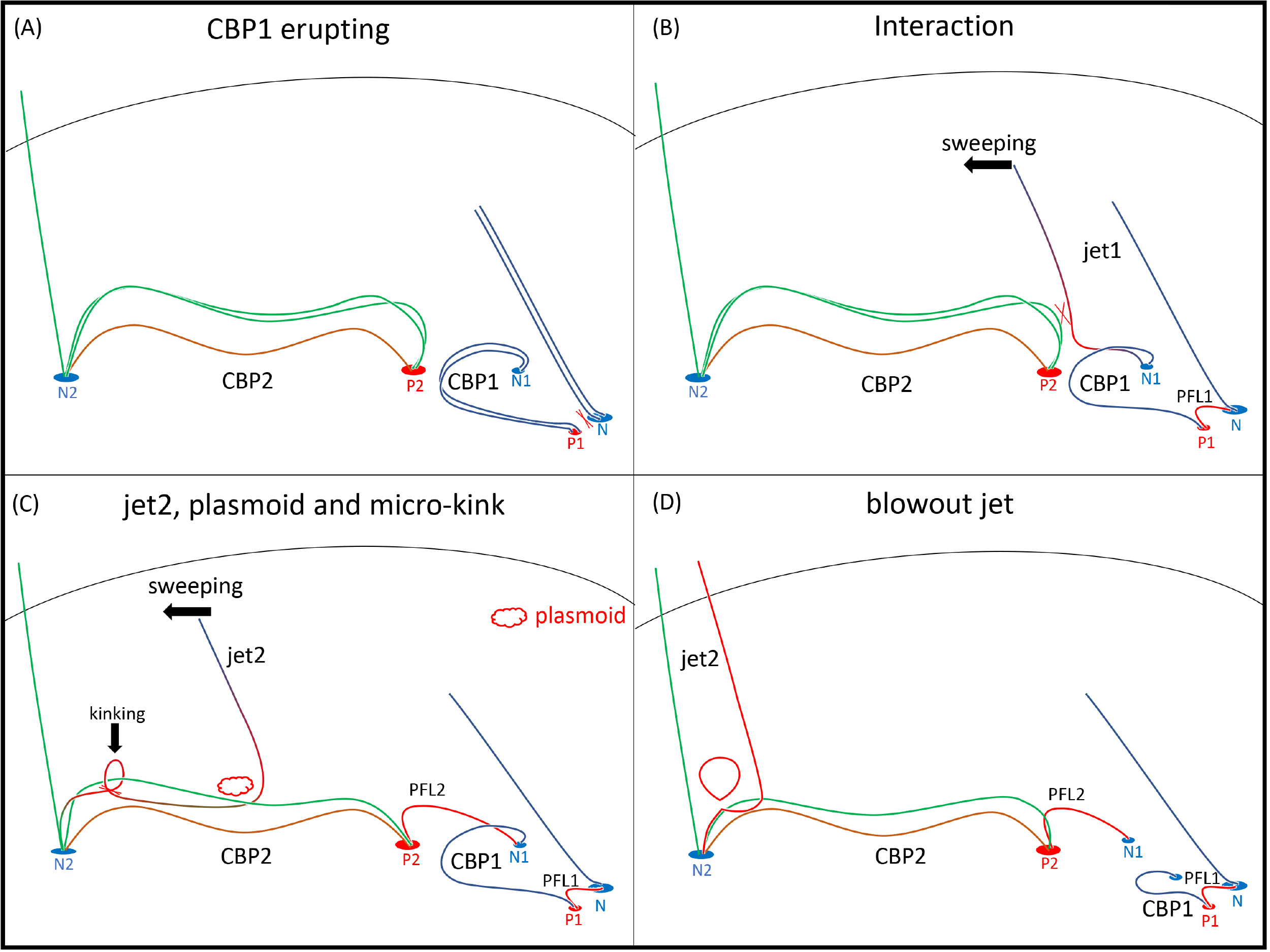}
\caption{A cartoon interpretation of the present event. Panel (A) shows the pre-eruption magnetic configuration; panel (B) shows the eruption of jet1 and its interaction with CBP2; panel (C) shows the formation of jet2 and the small kink structure; panel (D) shows the eruption of the small kink structure and the final stage of jet2. The associated magnetic polarities (N, N1, N2, P1, and N2) are the same as observed in the HMI magnetogram. CBP1 is represented by the two blue curves, while CBP2 is represented by the orange and green curves in panel (A). The reconnection sites are marked with ``X" symbols in panel (A) (B), and (C), and the newly formed reconnected field lines are highlighted with red lines or curves. The westward arrows in panels (B) and (C) show the lateral moving direction of jet1 and jet2, respectively. The downward arrow in panel (C) indicates the newly formed small kink structure. 		 
\label{fig5}}
\end{figure*}

To better present our explanation for the observational results, we draw a cartoon in \nfig{fig5} to demonstrate the eruption process of the present sympathetic jet event. In each panel of \nfig{fig5},  only a few representative field lines are drawn, and the orientation is the same with the images  displayed in previous figures (i.e., solar south up, east right). The newly formed reconnected field lines are plotted as red lines, and the reconnection sites are indicated by red ``X" symbols. The pre-eruption magnetic topologies of CBP1 and CBP2 are drawn in \nfig{fig5}(A), and the magnetic polarities (N, P1, N1, P2, and N2) keep the same as evidenced in the HMI LOS magentograms. CBP1 connects P1 and N1, and the ambient field lines nearby P1 were rooted in a negative polarity N. Due to some reasons such as the emergence of P1 and N1, the field lines of CBP1 start to reconnect with the open field line rooted in N (see \nfig{fig5}(A) and (B)). This reconnection would produce the bright patch (PFL1) as observed in soft X-ray and EUV images, and the reconnected open field lines will move transversely due to the magnetic tension force (slingshot effect). CBP2 was composed of two separated groups of loops but they were both rooted in P2 and N2. In the cartoon, we use brown and green curves to represent the two groups of loop systems in CBP2. The transverse motion of the spire of jet1 directly impacted upon the east part of CBP2 and triggered the magnetic reconnection between the two magnetic system. This reconnection not only created PFL2 connecting N1 and P2 but also a curving open loop which formed the spire of jet2. The whole evolution of both CBP1 and CBP2, in fact, can be referred to the open-closed reconnecting case reported by \cite{1996PASJ...48..353Y}. Similarly, the curving open loop attempt to straighten due to the magnetic tension force, and therefore it also moved westwardly. In addition, the lower part of this loop would form kink structures as evidenced in observations, due to the relaxing of the loop system caused by the reconnection and the westward moving plasmoids along the loop (see \nfig{fig5}(B) and (C)); such plasmoid has been found to exist nearby the current sheet during the reconnection both in observations and simulations \citep[e.g.,][]{2013ApJ...764...87L,2016ApJ...827....4W,2019ApJ...885L..15K}. Due to the kinking of the newly formed open loop, magnetic reconnection can occur below the kink structure between its two crossed legs, and the result of this reconnection is the formation of an upward erupting blob and the bright cusp-like structure around the west end of CBP2 (see \nfig{fig5}(D)). Therefore, we propose that jet1 and jet2 were respectively standard and blowout jets \citep{2010ApJ...720..757M}, and they were connected by the interaction between the spire of jet1 and CBP2.

\section{Conclusion \& Discussion}
Using soft X-ray and EUV observations taken by the {\em Hinode} and the {\em SDO}, we studied, for the first time, a sympathetic jet event in detail. In this event, two coronal jets were launched successively from two adjacent CBPs, and our analysis results indicate that they were physically connected to each other, rather than a coincidence of time. The first jet was a standard coronal jet produced by the magnetic reconnection between CBP1 with its ambient open field lines. The magnetic fluxes variation within the eruption source region showed emergence (cancellation) pattern before (after) the start of jet1. Therefore, we propose that the first jet was possibly triggered by magnetic flux emergence, and the cancellation may reflect the submergence of PFL1 that had a cusp-like shape. The second jet was a blowout jet involving the eruption of non-potential magnetic field within its base. The initiation of the second jet was caused by the reconnection between the spire of the first jet and CBP2, because of the direct impingement of the former upon the latter resulting from the lateral sweeping motion of the first jet. The violent eruption of the second jet was caused by the eruption of a kink structure within the jet base, which was triggered by the kink instability in the kink structure and then the reconnection between its two crossed legs. Our finds may contribute to the understanding of the following three aspects. 

As mentioned in the introduction section, the basic physical scenario for the formation of coronal jets is that the bipolar, mixed with plasma, reconnects with ambient field under certain circumstances, and sometimes, involving the eruption of mini-filaments. Generally, the photospheric magnetic flux within the eruption source region of a jet often show a first emergence and then cancellation variation pattern, and the transition time is often coincidence with the start time of the jet \citep[e.g.,][]{2011ApJ...735L..43S,2012ApJ...745..164S,2017ApJ...845...94T,2019ApJ...871..220S}. Thus, the flux emergence and cancellation are generally considered as important triggering agents for solar jets. However, the specific physical processes that magnetic flux emergence and cancellation represent are still unclear \citep{2021RSPSA.47700217S}. In the present observation, the magnetic flux within the eruption source region of the first jet showed emergence (cancellation) before (after) the start of the jet. We therefore propose that the launching of the first jet was caused by the reconnection between emerging bipole and the ambient open fields, and the flux cancellation may represent the submerging of the closed reconnected loops. Besides magnetic activities observed in the photosphere, coronal jets can also be generated from some progenitors in the solar atmosphere, such as mini-filaments, CBPs, and mini-sigmoids \citep{2021RSPSA.47700217S}. Observationally, the triggering of vast majority of coronal jets are due to the internal magnetic activities or magnetohydrodynamics instabilities within the eruption source regions. \cite{2014ApJ...786..151S} reported a special coronal jet occurred at the boundary of an equatorial coronal hole, which was caused by the passing of large-scale coronal waves which pushed the open loop in the coronal hole to reconnect with a low-lying CBP. That is to say, coronal jets can also be triggered by external disturbances. In the present study, the initiation of the second jet was due to the impingement of the spire of the first jet upon CBP2, which confirms that external disturbances do play as a trigger agent for the generation of some coronal jets. 

Kink structure is pervasive in the solar atmosphere, for examples in filaments and coronal loops. In observations, the writhing and kinking motions are attributed to the releasing of magnetic twist in strongly twisted magnetic structures such as those evidenced in failed filament eruptions \citep[e.g.,][]{2003ApJ...595L.135J,2006ApJ...653..719A,2012ApJ...750...12S}. Rotating jets typically exhibit the releasing of magnetic twist from closed magnetic system into open magnetic structure through magnetic reconnection \citep{2011ApJ...735L..43S,2019MNRAS.490.3679W}. \cite{2013ApJ...769..134M} found that the closed bipolar magnetic field in the jet base has substantial twist not only in all blowout jets but also in many standard jets, although blowout jets involving mini-filament eruption often exhibit stronger rotation motion \citep{2010ApJ...720..757M,2012ApJ...745..164S}. Since many jets are caused by the eruption of twisted mini-filament, it is reasonable to conjecture that writhing and kinking motions should be common in solar jets. However, although the rotating motion of solar jets exhibit the existence of twisting property of the mini-filaments or closed loops in the jet base, specific examples of kinking eruption within the jet base have not yet been reported. In our present case, the inverted-$\gamma$ kink structure exhibited the twisting property of the jet base, and its kink instability.  
What still remains unclear is whether the apparent solenoidal structure represents the propagation of twisted magnetic fields or is a coincidence of line-of-sight or emission effects. If there is accumulated twist in CBP2's magnetic system, then the solenoidal/helical structure may represent twist propagation during the reconnection associated with the CBP2 jet eruption. In other words, under such conditions, it is difficult to imagine how pre-eruption twist in the western region of CBP2 would remain trapped in the closed flux region during the dynamic jet eruption. However, since the observed, inverted-$\gamma$ structure is a clear signature of twist and is associated with the transient solenoid both spatially and temporally (see video for more details), it is likely this does represent twisted fields from CBP2. Taking this and the weak bipolar field strengths into account, we proposed that there is a moderate amount of twist stored in CBP2 before the eruption. But what is the key to ceasing or slowing down the twist propagation during the CBP2 eruption?  We offer the following potential explanation. Due to the magnetic tension, the newly opened magnetic field lines resulting from the interaction between the spire of the first jet and CBP2 would sweep towards west. At the same time, some of twist component of the first jet is transferred into CBP2, observed as writhing or unwinding. This interaction (reconnection) between the first jet and the western extension of CBP2 temporarily increases the twist in the closed-field regions of CBP2; during this process the twist in the spire of the fist jet can transfer into the second jet since the sweeping spire of the first jet before being expelled was transformed into the spire of the second jet through magnetic reconnection. This phenomenon is reasonably well resolved in our current study by SDO. Furthermore, the system relaxation associated with the eruption of the second jet's twist flux may be slowed down because of the inertia of the plasma frozen into the magnetic field. Thus, the release of the second jet's twist may share the same timescale as the sweeping motion interaction resulting from the first jet (i.e., the duration of the sweeping spire of the second is about 5 minutes). That is, despite the twist stored in CBP2 attempting to spread out as quickly as possible at the very beginning of the reconnection, this twist may not fully escape into the adjacent open field in such short time as 5 minutes. In other words, there exists the opportunity and possibility that the sweeping motion and reconnection temporarily increases the twist in the CBP2 system which is then observed as the helical or solenoidal structure that transitions into the inverted-$\gamma$ as part of the second jet's reconnection and eruption.

High spatiotemporal resolution observation of solar jets and the eruption of mini-filaments in recent years suggest that these small-scale solar eruptive activities exhibit similar physical properties with large-scale, energetic solar eruptions, and this may hint a scale invariance of solar eruptions  \citep{2000ApJ...530.1071W,2010ApJ...718..981R,2012ApJ...750...12S,2019ApJ...883..104S,2021RSPSA.47700217S,2015Natur.523..437S}. For sympathetic solar eruptions, the present event also manifests some similarity with other large-scale events, especially, for magnetic interaction events in the solar atmosphere. For example, during the interaction of a jet with a group of remote coronal loops, \cite{2008ApJ...677..699J} found that the eruption resulted in two CMEs within two hours, in which one was associated with the jet, the other was associated with the eruption of the loops. In our event, due to the nature of the observed jets (i.e., their pre-eruption configuration and open-closed reconnection), there was no large-scale CMEs caused by the eruptions. One can found that no matter small-scale jets or large-scale solar eruptions, an eruption can cause the launching of another one at different location in a suitable magnetic environment. 

The authors would like to thank the helpful discussions with Dr.Yang Liheng from Yunnan Observatories, and also the anonymous referee for his/her many valuable suggestions and comments for improving the quality of this paper. Moreover, the authors want to acknowledge {\em SDO}/AIA and {\em SDO}/HMI, and {\em Hinode} science teams for providing the data. {\em Hinode} is a Japanese mission developed and launched by ISAS/JAXA, with NAOJ as a domestic partner and NASA and STFC (UK) as international partners, which is operated by these agencies in co-operation with ESA and NSC (Norway). This work is supported by the National Key R\&D Program of China (2019YFA0405000), the Natural Science Foundation of China (11922307,11773068,11633008), the Yunnan Science Foundation (2017FB006), the Specialized Research Fund for State Key Laboratories, and the West Light Foundation of Chinese Academy of Sciences. The research by A. Elmhamdi was supported by King Saud University, Deanship of Scientific Research, College of Science Research Center.

\end{document}